\documentclass[preprint,amsmath,amssymb,aps]{revtex4}

\usepackage{graphicx}
\usepackage{xcolor}

\begin{document}

\title{Cosmological model with local symmetry of very special relativity and constraints on it from supernovae}

\author{Zhe Chang$^{1,2}$\footnote{E-mail: changz@ihep.ac.cn}}
\author{Ming-Hua Li$^{1}$\footnote{E-mail: limh@ihep.ac.cn}}
\author{Xin Li$^{1,2}$\footnote{E-mail: lixin@ihep.ac.cn}}
\author{Sai Wang$^{1}$\footnote{E-mail: wangsai@ihep.ac.cn}\footnote{Corresponding author at IHEP, CAS, 100049 Beijing, China.}}
\affiliation{${}^1$Institute of High Energy Physics\\ Chinese Academy of Sciences, 100049 Beijing, China\\
${}^2$Theoretical Physics Center for Science Facilities\\ Chinese Academy of Sciences, 100049 Beijing, China}

\begin{pacs}
{98.80.Es, 98.80.Jk}
\end{pacs}

\begin{abstract}
Based on Cohen \& Glashow's very special relativity [A. G. Cohen and S. L. Glashow, Phys. Rev. Lett. {\bf 97} (2006) 021601], we propose an anisotropic modification to the Friedmann-Robertson-Walker (FRW) line element. An arbitrarily oriented 1-form is introduced and the FRW spacetime becomes of the Randers-Finsler type. The 1-form picks out a privileged axis in the universe. Thus, the cosmological redshift as well as the Hubble diagram of the type Ia supernovae (SNe Ia) becomes anisotropic. By directly analyzing the Union2 compilation, we obtain the privileged axis pointing to \((l,b)=({304^\circ}\pm{43^\circ},{-27^\circ}\pm{13^\circ})\) (\(68\%~\rm{C.L.}\)). This privileged axis is close to those obtained by comparing the best-fit Hubble diagrams in pairs of hemispheres. It should be noticed that the result is consistent with isotropy at the \(1\sigma\) level since the anisotropic magnitude is \(D=0.03\pm 0.03\).
\end{abstract}
\maketitle

\section{Introduction}
The cosmological principle is one of foundations of the standard cosmological model, i.e., the \(\Lambda\)CDM model \cite{Cosmology by Weinberg}.
It says that the universe is statistically homogeneous and isotropic at large scale.
The \(\Lambda\)CDM model is well consistent with present cosmological observations,
such as the Wilkinson Microwave Anisotropy Probe (WMAP) \cite{WMAP7} and the Sloan Digital Sky Survey (SDSS) \cite{SDSS}, etc.
However, there exist challenges for the standard cosmological model (see review in Ref.~\cite{LCDM: Triumphs Puzzles and Remedies}),
such as the large-scale cosmic flows \cite{Consistently large cosmic flows on scales of 100 Mpc: a challenge for the standard CDM cosmology},
the alignment of low multipoles in the CMB spectra \cite{Seven-year Wilkinson Microwave Anisotropy Probe (WMAP) Observations: Are There Cosmic Microwave Background Anomalies,The axis of evil,High resolution foreground cleaned CMB map from WMAP},
the large-scale alignment of the quasar polarization vectors \cite{Mapping extreme-scale alignments of quasar polarization vectors}.
One of their resolutions refers to a privileged axis at large scale
in the universe \cite{Searching for a cosmological preferred axis: Union2 data analysis and comparison with other probes}.

The type Ia supernovae (SNe Ia) have been used to search for the possible anisotropy of the universe \cite{Constraints on cosmological anisotropy,Dipole of the Luminosity Distance,Cosmological Constraints from Type Ia Supernovae Peculiar Velocity Measurements,(An)isotropy of the Hubble diagram: comparing hemispheres,Direction dependence and non-Gaussianity in the high-redshift supernova data,Anisotropic Dark Energy,Constraining dark energy fluctuations with supernova correlations,Measuring dark energy spatial inhomogeneity with supernova data,Direction dependence in supernova data: constraining isotropy,On the Possibility of Anisotropic Curvature in Cosmology,Probing the anisotropic local universe and beyond with SNe Ia data,Testing the isotropy of the universe with type Ia supernovae}, since they were employed to discover the cosmic acceleration \cite{Cosmic acceleration by Riess,Cosmic acceleration by Perlmutter}.
Especially, Antoniou \& Perivolaropoulos \cite{Searching for a cosmological preferred axis: Union2 data analysis and comparison with other probes}
used the hemisphere comparison method to analyze the Union2 data \cite{Union2} and
found a direction \((l,b)=({309^\circ}^{+23^\circ}_{-3^\circ},{18^\circ}^{+11^\circ}_{-10^\circ})\)
for the maximum accelerating expansion of the universe.
Similarly, Cai \& Tuo \cite{Direction Dependence of the Deceleration Parameter} found a preferred direction \((l,b)=({314^{\circ}}^{+20^{\circ}}_{-13^{\circ}},{28^{\circ}}^{+11^{\circ}}_{-33^{\circ}})\) for the cosmological deceleration parameter.
Most recently, Kalus {\it et al.} \cite{Constraints on anisotropic cosmic expansion from supernovae} also found that the highest expansion rate of the universe towards the direction \((l,b)\approx(-35^{\circ},-19^{\circ})\) (\(95~\%~\rm{C.L.}\)).

A privileged axis, as is mentioned above, may account for the anisotropic phenomenologies of the observational astrophysics and cosmology.
Actually, the issue of privileged axis has been studied extensively in the very special relativity (VSR) \cite{VSR}.
The Finslerian spacetime structure \(d\tau=\left(\eta_{\mu\nu}dx^{\mu}dx^{\nu}\right)^{\frac{1-b}{2}}\left(n_{\sigma} dx^{\sigma}\right)^{b}\)
was proved to be invariant under the \(DISIM_{b}(2)\) group \cite{VSR in Finsler}.
There is a preferred axis \(n_{\sigma}\) in the above line element.
It could characterize the anisotropy of the flat spacetime, which leads to the Lorentz invariance violation (LIV).
Locally, the Randers metric was proved to possess the symmetry of the group \(TE(2)\) \cite{Finsler isometry LiCM,The Finsler Type of Space-time Realization of Deformed Very Special Relativity}.
The group \(TE(2)\) is a semi-product of \(T(4)\) and \(E(2)\) \cite{The Deformation of Poincar¨¦ Subgroups Concerning Very Special Relativity}.
The group \(E(2)\) denotes a proper subgroup of the Lorentz group with three generators \cite{VSR}.
Therefore, Randers spacetime could possess local symmetry of the generic VSR.

In this paper, we propose a Randers line element (structure) \cite{Randers space}
with local symmetry of the generic VSR to estimate the possible anisotropy of the Hubble diagram of SNe Ia.
At first order, we show the related Friedmann equation which is same to that in the standard model.
A modified cosmological redshift formula is presented.
The redshift is direction dependent and refers a privileged axis. This may imply
that the universe undergoes an anisotropic expansion. Next, we show
a modified luminosity-distance vs. redshift relation for the SNe Ia. It is certainly
anisotropic. The Union2 compilation \cite{Union2} is used to
constrain the direction of the privileged axis.
The rest of the paper is arranged as follows.
In section 2, the Friedmann equation is presented in the Randers spacetime.
The anisotropic Hubble diagram is showed at first order in section 3.
In section 4, we investigate the Union2 dataset of the supernovae to constrain the level of the anisotropy of the universe.
Conclusions and discussions are listed in section 5.

\section{The Friedmann equation in Randers spacetime}

In the \(\Lambda\)CDM model, the cosmic spacetime is described by the spatially flat Friedmann-Robertson-Walker (FRW)
line element \cite{Cosmology by Weinberg}
\begin{equation}
\label{FRW metric}
\overline{d\tau}=\sqrt{dt^2-a^2(t)\left(dx^2+dy^2+dz^2\right)} \ ,
\end{equation}
where $a(t)$ denotes the scale factor of the universe at the time \(t\).
The cosmological redshift \(\overline{z}\) is given by
\begin{equation}
\label{redshiftinFriemann}
1+\overline{z}(t)=\frac{a(t_{0})}{a(t)}\equiv\frac{1}{a(t)}\ ,
\end{equation}
where we set \(a(t_{0})\equiv 1\) for today.
The redshift describes the expansion rate of the universe between \(t\) and \(t_{0}\).
The overline denotes physical objects in the FRW-Riemannian spacetime.

The FRW line element should be modified when there exists a privileged axis in the universe.
We postulate that an extra 1-form is added into the FRW structure.
The 1-form singles out a privileged axis in the universe.
The spacetime structure becomes
\begin{equation}
\label{Randers-FRW}
d\tau\equiv\overline{d\tau}+\tilde{b}_{\mu}(x)dx^{\mu}\ .
\end{equation}
Actually, the spacetime structure (\ref{Randers-FRW}) belongs to Randers type \cite{Randers space}.
This is a class of Finsler spacetime \cite{Book by Bao}.
The 1-form could be viewed as an arbitrarily oriented electromagnetic 4-potential in the universe.
It may be the relic of a primordial magnetic field at large scales \cite{Primordial magnetic field 01,Primordial magnetic field 02,Primordial magnetic field 03,Primordial magnetic field Evidence}.

In the following sections, we will show that the anisotropy of Hubble diagram stems from
the spatial components (3-vector) of the preferred axis \(\tilde{b}_{\mu}\).
In addition, we can choose the coordinate system such that the spatial 3-vector is the third spatial axis.
In this way, the 1-form becomes \(\tilde{b}_{\mu}dx^{\mu}=\tilde{b}_{3}(t)dz\).
Here \(\tilde{b}_{3}\) is set to evolve with only \(t\).
Without loss of generality, thus, the Randers structure (\ref{Randers-FRW}) could be simplified as
\begin{equation}
\label{Randers structure b3}
d\tau=\sqrt{dt^2-a^2(t)\left(dx^2+dy^2+dz^2\right)}+\tilde{b}_{3}(t)dz\ .
\end{equation}
Following the Stavrinos {\it et al.}'s approach \cite{FRW model with weak anisotropy by Stavrinos},
the above Randers structure could be approximated as one osculating Riemann metric
via the osculating Riemannian method \cite{Book by Rund}.
The Randers spacetime evolves following along the conventional
Einstein's gravitational field equations in this approach \cite{Book by Asanov}.

In the osculating Riemannian method, we could view the velocity coordinates as functions of the position, i.e. \(y=y(x)\),
by restricting the velocity \(y^{\mu}\equiv \frac{dx^{\mu}}{d\tau}\) to an individual tangent space for the given position \(x^{\mu}\).
In this way, the Finsler geometric quantities are approximated by the corresponding osculating Riemannian quantities.
The Einstein's gravitational field equations could be obtained by computing the connection and the curvature of the osculating Riemannian metric \(g_{\mu\nu}(x)\equiv g_{\mu\nu}(x,y(x))\), see details in the references \cite{FRW model with weak anisotropy by Stavrinos,Book by Rund,Book by Asanov,F-Inflation}.
After a calculation of the time-time component of Einstein's field equations,
one obtains the Friedmann equation in the Randers spacetime.
The Friedmann equation is given by
\begin{equation}
\label{Friedmann equation}
3\left(\frac{\dot{a}}{a}\right)^{2}+\frac{1}{a^{4}}\left(\left(\dot{a}^{2}-2a\ddot{a}\right)\tilde{b}_{3}^{2}
-a\dot{a}\left(\tilde{b}_{3}^{2}\right)^{\cdot}\right)=8\pi G \rho\ ,
\end{equation}
where the dots denote the temporal derivative $\frac{d}{dx^0}$ and \(\rho\) is the energy density of cosmic inventory.
We notice that the second term on the left hand side involves only the second-order effects,
which are proportional to \(\tilde{b}_{3}^{2}\) and its temporal derivative.
This term could be dropped at the first-order approximation.
Thus, the Friedmann equation (\ref{Friedmann equation}) reduces back to
the conventional Friedmann equation as
\begin{equation}
\left(\frac{\dot{a}}{a}\right)^{2}=\frac{8\pi G}{3}\rho\ .
\end{equation}
This result reveals that the anisotropic effect is secondary for the dynamical evolution of
the Randers spacetime (\ref{Randers structure b3}).
However, the anisotropic effect could have significant influence on the kinematical behaviors of the Randers spacetime.
We will discuss this issue in the following section.

\section{The anisotropic Hubble diagram}

In the FRW-Randers spacetime (\ref{Randers-FRW}), the cosmological redshift \(z\) has
been derived via resolving the Finslerian geodesic equations, which are given by \cite{Fine
structure constant variation or spacetime anisotropy}
\begin{eqnarray}
\label{geodesic eq1}
\frac{d^2x^0}{d\tau^2}+\delta_{ij}\dot{a}a\frac{dx^i}{d\tau}\frac{dx^j}{d\tau}+\frac{d x^{0}}{d\tau}f\left(x,\frac{dx}{d\tau}\right)&=&0\ ,\\
\label{geodesic eq2}
\frac{d^2x^i}{d\tau^2}+2\delta^i_j\frac{\dot{a}}{a}\frac{dx^0}{d\tau}\frac{dx^j}{d\tau}+\frac{dx^i}{d\tau}f\left(x,\frac{dx}{d\tau}\right)&=&0\ ,
\end{eqnarray}
where \(f\left(x,\frac{dx}{d\tau}\right)\equiv\tilde{b}_{\nu|\lambda}\frac{dx^\nu}{d\tau}\frac{dx^\lambda}{d\tau}/F\).
Here \(\tilde{b}_{\mu|\nu}\) denotes the covariant derivative of \(\tilde{b}(x)\)
with respect to the FRW metric \cite{Book by Bao}.
The above geodesic equations have solutions as \(a\frac{dx^0}{d\tau}\propto J_1\) and \(a^2\frac{dx^i}{d\tau}\propto J_1\),
where \(J_1=1-\tilde{b}_{\mu}\widehat{p}^{\mu}\) \cite{Fine
structure constant variation or spacetime anisotropy}.
Thus, the redshift \(z\) could be written as
\begin{equation}
\label{Randers redshift}
1+z(t,\widehat{p})=\frac{1}{a(t)}\left(1-\tilde{b}_{\mu}\widehat{p}^{\mu}\right)\ ,
\end{equation}
where the unit 4-vector $\widehat{p}$ denotes a light-like direction towards each SN Ia.
It is anisotropic since there is a privileged axis \(\tilde{b}_{\mu}(x)\) in the above formula.

When the privileged axis is a spatial 3-vector, the modified redshift \(z\) could be rewritten as
\begin{equation}
1+z(t,\widehat{\textbf{p}})=\frac{1}{a(t)}\left[1-D(\widehat{\textbf{n}}\cdot\widehat{\textbf{p}})\right]\ ,
\end{equation}
where \(\widehat{\textbf{n}}\) is the unit 3-vector for the spatial components of \(\tilde{b}\),
and \(D\) denotes the magnitude which is smaller than one.
To simplify our following discussions, we choose the spatial 3-vector \(\widehat{\textbf{n}}\) as the third spatial axis of the coordinate system.
Thus, the redshift \(z\) becomes
\begin{equation}
\label{redshift}
1+z(t,\cos\theta)=\frac{1}{a}\left[1-D\cos\theta\right]\ ,
\end{equation}
where \(\theta\) denotes the angle between \(\widehat{\textbf{n}}\) and
\(\widehat{\textbf{p}}\). This formula shows clearly that the
cosmological redshift is direction dependent and anisotropic.
Thus, the light undergoes an anisotropic propagation of the spatial dipole form in the universe.

The Hubble diagram would be anisotropic and have a
corresponded privileged axis which exists in the Randers
spacetime. We will discuss this proposition in the following
paragraphs. For the null geodesic, the spacetime line element
vanishes, i.e., \(d\tau=0\). In the FRW-Randers spacetime, the Randers
structure (\ref{Randers-FRW}) should vanishes. Thus, we obtain
\begin{equation}
dt^{2}-a^{2}dr^{2}=\left(-D\cos\theta dr\right)^{2}\ ,
\end{equation}
where we use the polar coordinates and \(\widehat{\textbf{n}}\) is set
again as the third spatial axis. Here the angular coordinates are
discarded since they are irrelative to the distance-redshift relation.
Finally, the above equation could be simplified as
\begin{equation}
\label{dt-dr}
dt=\sqrt{a^{2}+D^{2}\cos^{2}\theta} dr\ .
\end{equation}
The privileged axis emerges again, which affects the propagation of the cosmic light.
This result is different from that of \(dt=a(t)dr\) in the \(\Lambda\)CDM model.
The modification is quadrupolar in Eq.~(\ref{dt-dr}).
Obviously, the cosmic light would propagate with different speeds in different directions in the space.
However, the quadrupolar effect is a second-order effect.
It just affects the propagation of the cosmic light slightly.
Thus, we could disregard it in the following discussion and the equation (\ref{dt-dr}) becomes the conventional one \(dt=a(t)dr\).

In the observational universe, the luminosity distance \(d_{L}\) of a SN Ia is given by \cite{Book by Dodelson}
\begin{equation}
d_{L}\equiv(1+z) r\ ,
\end{equation}
where \(r\) denotes the comoving distance between us and the SN Ia.
We could obtain the comoving distance \(r\) by integrating the equation (\ref{dt-dr}) at the first order of \(D\cos\theta\).
Therefore, the luminosity-distance vs. redshift relation can be obtained as
\begin{equation}
\label{distance-redshift relation}
d_{L}=(1+z)\int^{t_{0}}_{t(z)}\frac{dt'}{a(t')}\ ,
\end{equation}
which is same to the conventional one.
It could be rewritten as
\begin{equation}
\label{luminosity-distance and redshift relation}
d_{L}=(1+z)\int^{1}_{(1+z)^{-1}}\frac{da}{a^2H_0\sqrt{\Omega_{m}a^{-3}+\Omega_\Lambda}}\ ,
\end{equation}
where \(\Omega_{m}\equiv(8\pi G/3H_0^2)\rho_{m0}\) and \(\Omega_\Lambda\equiv\Lambda/3H_0^2\), \(\rho_{m0}\)
is the critical mass density, $\Lambda$ is the cosmological constant, and $H_0$ is the Hubble constant.
Here we assume that the energy density of cosmic inventory is critical today, i.e., \(\Omega_{m}+\Omega_{\Lambda}=1\).
In the derivation of (\ref{luminosity-distance and redshift relation}), we have used the conventional Friedmann equation
and the definition $H\equiv \dot{a}/a$ as a first-order approximation.

By substituting the redshift (\ref{redshift}) into the equation (\ref{luminosity-distance and redshift relation}),
we obtain the anisotropic distance-redshift relation as
\begin{equation}
\label{Anisotropic Hubble diagram}
H_{0}d_{L}=(1+z)\int_{0}^{z}\frac{\mathcal{A}^{-1}dz'}{\sqrt{\Omega_{m}\left(\frac{\mathcal{A}}{1+z'}\right)^{-3}+\Omega_{\Lambda}}}\ ,
\end{equation}
where \(\mathcal{A}\equiv 1-D\cos\theta\).
At first order, the above equation becomes
\begin{equation}
\label{Anisotropic Hubble diagram first order}
H_{0}d_{L}=(1+z)\int_{0}^{z}\frac{\mathcal{B}(z',\cos\theta)dz'}{\sqrt{\Omega_{m}(1+z')^{-3}+\Omega_{\Lambda}}}\ ,
\end{equation}
where \(\mathcal{B}\) denotes the dipolar effect,
\begin{equation}
\label{B}
\mathcal{B}(z,\cos\theta)\equiv 1+\frac{-{1}/{2}+\Omega_{\Lambda}/\Omega_{m}(1+z)^{3}}{1+\Omega_{\Lambda}/\Omega_{m}(1+z)^{3}}D\cos\theta\ .
\end{equation}
These relations refer to the anisotropic Hubble diagram of SNe Ia.
The direction dependence is obvious in the above equations
since different modifications are introduced to the distance-redshift relation of SNe Ia towards different spatial directions.
Once again, we see that the 1-form in the Randers structure leads to the privileged axis in the universe.
When \(D\) vanishes, both \(\mathcal{A}\) and \(\mathcal{B}\) reduce back to \(1\).
Thus, the FRW-Randers spacetime returns back to the \(\Lambda\)CDM model.

\section{Numerical results from supernovae}
We define a Cartesian coordinate system for the unit 3-vector \(\widehat{\textbf{p}}\) corresponding to each SN Ia
with the equatorial coordinates $(\alpha, \delta)$, in which \(\widehat{\textbf{p}}\) is given as
\begin{equation}
\label{pi}
\widehat{\textbf{p}}\equiv \cos(\delta) \cos(\alpha)~\hat{\textbf{i}}
+ \cos(\delta)\sin(\alpha) ~\hat{\textbf{j}} + \sin(\delta)~ \hat{\textbf{k}}\ ,
\end{equation}
where \(\hat{\textbf{i}}\), \(\hat{\textbf{j}}\) and \(\hat{\textbf{k}}\) are basis 3-vectors.
Suppose \(\widehat{\textbf{n}}\) is given by
\begin{equation}
\label{n}
\widehat{\textbf{n}}\equiv \cos(\delta_0)
\cos(\alpha_0)~\hat{\textbf{i}} + \cos(\delta_0)~\sin(\alpha_0) ~\hat{\textbf{j}} +
\sin(\delta_0)~ \hat{\textbf{k}}\ ,
\end{equation}
then the cosine of the angle $\theta$ between these two vectors as
\begin{equation}
\label{theta} \cos\theta \equiv \widehat{\textbf{n}} \cdot
\widehat{\textbf{p}}\ .
\end{equation}
Substituting the relation (\ref{theta}) into (\ref{B}),
the theoretical distance modulus could be calculated from (\ref{Anisotropic Hubble diagram first order}).
It is given as
\begin{equation}
\label{distance modulus}
\mu_{\rm th}(z, \alpha, \delta; \alpha_0, \delta_0, D)=5 \textmd{log}_{10} [d_L(\textmd{Mpc})] + 25\ .
\end{equation}
Here \(\mu_{th}\) has the direction dependence and the privileged axis points
towards the direction \((\alpha_0,\delta_0)\) in the equatorial coordinate system.

We base our numerical study on the Union2 dataset \cite{Union2}. The Union2 compilation consists of 557 SNe Ia.
Thus, we perform the least-$\chi ^2$ fit to the data of these SNe Ia to determine the privileged axis $(\alpha_0, \delta_0)$
and the geometrical parameter \(D\) (viewed as a constant for simplicity).
The $\chi^2$ statistic in our fit is
\begin{eqnarray}
\label{chisquare}
\chi^2 \equiv \sum_{i=1}^{557}{[
\mu_{\rm th}(z_i, \alpha_i, \delta_i; \alpha_0, \delta_0, D)-\mu_{\rm obs}(z_i)]^2 \over {{\sigma_{\mu}(z_i)}^2}}\ ,
\end{eqnarray}
where $\mu_{\rm th}(z_i, \alpha_i, \delta_i; \alpha_0, \delta_0, D)$ is the theoretical distance modulus. $\mu_{\rm
obs}(z_i)$ and $\sigma_{\mu}(z_i)$, respectively, denote the
observational values of the distance modulus and measurement errors,
which are obtained from the Union2 compilation.
In the FRW-Randers spacetime, \(D\) is at the level of \(\sim0.01\),
which leads to only first-order term to the equation (\ref{Anisotropic Hubble diagram first order}).

\begin{figure*}[h]
\begin{center}
\includegraphics[width=6 cm]{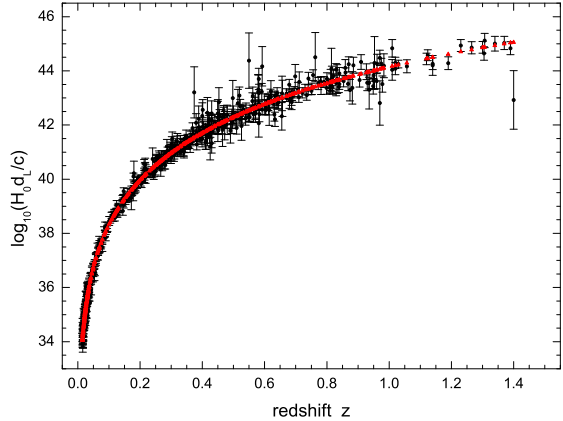}
\caption{The distance modulus vs. redshift relation of the SNe Ia in the FRW-Randers spacetime. The (black) dots and errorbars denote experimental data which comes from Union2 compilation \cite{Union2}. The (red) triangles denote theoretical predictions for the SNe Ia in our model.}
\label{AnisotropicSNeIa}
\end{center}
\end{figure*}

\begin{figure*}[h]
\begin{center}
\includegraphics[width=9 cm]{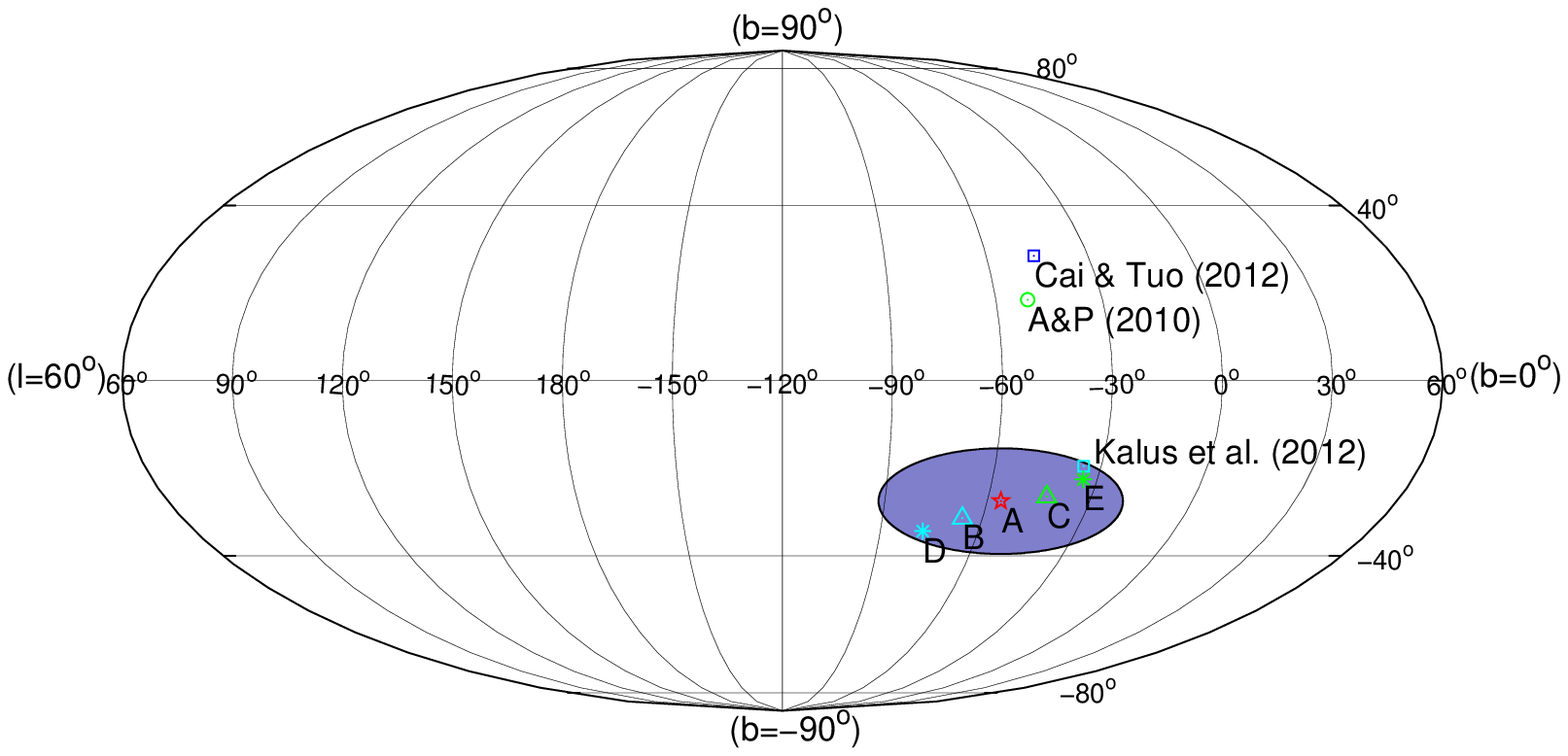}
\caption{The direction of privileged axis as well as its possible variation in the galactic coordinate.
The point A denotes the direction of privileged axis obtained in this paper $(l,b)=({304^\circ}\pm{43^\circ},{-27^\circ}\pm{13^\circ})~(68\%~\rm{C.L.})$
with the fixed parameters \(\Omega_m=0.27\) and \(H_0=70.0\) as the center values.
For the \(1\sigma\)-error variation of \(\Omega_m\) (\(H_0=70.0\)), the privileged axis points to the direction
\((l,b)=(294^\circ,-31^\circ)\) (point B) with \(\Omega_m=0.29\),
and to \((l,b)=(317^\circ,-26^\circ)\) (point C) with \(\Omega_m=0.25\).
For the \(1\sigma\)-error variation of \(H_0\) (\(\Omega_m=0.27\)),
it points to the direction \((l,b)=(283^\circ,-34^\circ)\) (point D) with \(H_0=70.4\),
and to \((l,b)=(326^\circ,-22^\circ)\) (point E) with \(H_0=69.6\).
For contrast, we also show other possible directions
of the privileged axis obtained via the hemisphere comparison method \cite{Searching for a cosmological
preferred axis: Union2 data analysis and comparison with other
probes,Direction Dependence of the Deceleration Parameter,Constraints on anisotropic cosmic expansion from supernovae}.}
\label{preferreddirection}
\end{center}
\end{figure*}

We first carry the least-$\chi^2$ fit of the $\Lambda$-CDM model to the Union2 compilation.
We obtain the best-fit parameters, i.e., the matter component \(\Omega_{m}=0.27\pm 0.02\),
and the Hubble parameter \(H_0=70.0\pm 0.4\) (the unit is \(\rm{km\cdot s^{-1}\cdot Mpc^{-1}}\) throughout the paper).
Note that these fitted parameters are consistent with those obtained by the WMAP \cite{WMAP7}.
In the following, the parameters \(\Omega_m\) and \(H_0\) are fixed to the above values as the center values.
Then we employ the least-\(\chi^2\) method (\ref{chisquare}) to obtain the anisotropic magnitude \(D\)
and the direction of privileged axis \((\alpha_0,\delta_0)\).
The results are given as $D=0.03\pm0.03$,
and \((\alpha_0,\delta_0)=({263^\circ}\pm{43^\circ},-{89^\circ}\pm{13^\circ})\) in the equatorial coordinate system
or \((l,b)=({304^\circ}\pm{43^\circ},{-27^\circ}\pm{13^\circ})\) in
the galactic coordinate system, for a minimum value of \(\overline{\chi^2}_{\textmd{min}}\simeq 0.97\).
All the results are presented in a sense of $68\%$ C.L..
Correspondingly, the distance modulus vs. redshift relation of the SNe Ia is shown in Fig.~\ref{AnisotropicSNeIa}.
The direction (point A) of privileged axis obtained by this direct analysis is close to those obtained by comparing the best-fit Hubble diagrams in pairs of hemispheres,
see Fig.~\ref{preferreddirection}.
For example,
Antoniou \& Perivolaropoulos \cite{Searching for a cosmological
preferred axis: Union2 data analysis and comparison with other
probes} showed
\((l,b)=({309^\circ}^{+23^\circ}_{-3^\circ},{18^\circ}^{+11^\circ}_{-10^\circ})\),
Cai \& Tuo \cite{Direction Dependence of the Deceleration Parameter}
got \((l,b)=({314^{\circ}}^{+20^{\circ}}_{-13^{\circ}},{28^{\circ}}^{+11^{\circ}}_{-33^{\circ}})\),
and Kalus {\it et al.} \cite{Constraints on anisotropic cosmic expansion from supernovae} obtained
\((l,b)\approx(-35^{\circ},-19^{\circ})\) (\(95~\%~\rm{C.L.}\)).

We have fixed the parameters \(\Omega_m\) and \(H_0\) in the above fit to find the privileged axis.
Related to these center values, the direction of privileged axis is denoted
by the point A in Fig.~\ref{preferreddirection} as is mentioned above.
To see the possible variation of the direction of privileged axis, we vary respectively the \(\Omega_{m}\) and
\(H_0\) by $1\sigma$ error with respect to their center values, i.e., \(\Omega_m=0.27\) and \(H_0=70.0\).
For the \(1\sigma\)-error variation of \(\Omega_m\) (\(H_0=70.0\)), the privileged axis points towards to
the direction \((l,b)=(294^\circ,-31^\circ)\) (point B) with \(\Omega_m=0.29\),
and to \((l,b)=(317^\circ,-26^\circ)\) (point C) with \(\Omega_m=0.25\).
For the \(1\sigma\)-error variation of \(H_0\) (\(\Omega_m=0.27\)),
it points to the direction \((l,b)=(283^\circ,-34^\circ)\) (point D)
with \(H_0=70.4\), and to \((l,b)=(326^\circ,-22^\circ)\) (point E) with \(H_0=69.6\).
The above obtained directions are also depicted in Fig.~\ref{preferreddirection}.
We could see that all these directions are close to each other.
Thus, the above results reveal that the direction of privileged axis remains almost same,
when the cosmological parameters \(\Omega_m\) and \(H_0\) are allowed to vary by \(1\sigma\) error, respectively.

\section{Conclusions and remarks}
In this paper, we proposed an anisotropic modification to the FRW line element.
The modified line element refers to the Randers spacetime, which possesses the local symmetry of Cohen \& Glashow's VSR.
The local symmetry involves the group \(TE(2)\).
The Euclidean group \(E(2)\) contains three generators \(T_{1}\equiv K_{x}+J_{y}\), \(T_{2}\equiv K_{y}-J_{x}\), and \(J_{z}\),
where \(K_{i}\) and \(J_{i}\) (\(i=x,y,z\)) denote the generators of boosts and rotations, respectively.
The former two generators form a two-parameter group of translations in the \(x-y\) plane.
Thus, the local FRW-Randers spacetime is cylindrically symmetric.
However, the parity violates in the \(z\)-direction.
Otherwise, the \(E(2)\) would be enlarged to the Lorentz group \cite{VSR}.
Actually, the 1-form \(\tilde{b}_{\mu}dx^{\mu}\) in the Randers structure changes its sign
under the direction reversal \({dx^{\mu}}/{d\tau}\longrightarrow-{dx^{\mu}}/{d\tau}\).
This reveals that the Randers structure is asymmetric.
On the other hand, the indicatrix of the FRW-Randers spacetime is quadratic,
while the center of the indicatrix is displaced in the \(z\)-direction \cite{Randers space}.
The above two accounts imply a privileged axis in the FRW-Randers spacetime.

The existence of the privileged axis contradicts with the cosmological principle and implies a statistically anisotropic universe.
We extracted the direction \((l,b)=({304^\circ}\pm{43^\circ},{-27^\circ}\pm{13^\circ})\) (\(68\%~\rm{C.L.}\)) for the privileged axis \(\tilde{b}^{i}\), based on the Union2 compilation of the SNe Ia.
This direction is close to those obtained by comparing the best-fit Hubble diagrams in pairs of hemispheres.
It is noteworthy that the preferred direction coincides within its error bars with the direction \(\hat{d}_1\)
of partial mirror antisymmetry of CMB temperature fluctuations \cite{Searching for hidden mirror symmetries in CMB fluctuations from WMAP 7 year maps}.
In addition, one should note that our result is consistent with isotropy at the \(1\sigma\) level.
The reason is that the anisotropic magnitude was obtained as \(D=0.03\pm 0.03\) in our best-fit.

It is noteworthy that the Randers spacetime belongs to Finsler geometry \cite{Book by Bao}.
Actually, Finsler geometry gets rid of the quadratic restriction on the spacetime structure (line element) \cite{Chern}.
It is a natural generalization of Riemann geometry and includes Riemann geometry as a special case.
Finsler spacetime could depend on certain preferred directions
of the spacetime background \cite{VSR in Finsler,A special-relativistic theory of the locally anisotropic space-time. I,A special-relativistic theory of the locally anisotropic space-time. II,A special-relativistic theory of the locally anisotropic space-time. Appendix,Electromagnetic field in Finsler,Constraints on spacetime anisotropy and Lorentz violation from the GRAAL experiment,Geometrical Models of the Locally Anisotropic Space-Time}.
This could also be revealed via the isometric transformation \cite{Finsler isometry LiCM,Finsler isometry by Wang,Finsler isometry by Rutz}.
There are no more than \(\left(\frac{d(d-1)}{2}+1\right)\) Killing vectors in the d-dimensional Finsler spacetime \cite{Finsler isometry by Wang}.
Otherwise, Finsler spacetime becomes Riemannian.
Thus, it might be a reasonable candidate to account for the privileged axis and the anisotropic properties of the universe.

\vspace{0.2 cm}
\begin{acknowledgments}
We thank useful discussions with Jian-Ping Dai, Yunguo Jiang, Danning Li, and Hai-Nan Lin.
We are very grateful to Prof. Shuang-Nan Zhang who provides us the galactic coordinates of SNe Ia in the Union2 compilation.
This work is supported by the National Natural Science Fund of China under Grant No. 11075166.
\end{acknowledgments}

\end{document}